\documentclass{article}
\usepackage{arxiv}

\usepackage[utf8]{inputenc} 
\usepackage[T1]{fontenc}    
\usepackage{hyperref}       
\usepackage{url}            
\usepackage{booktabs}       
\usepackage{amsfonts}       
\usepackage{nicefrac}       
\usepackage{microtype}      
\usepackage{lipsum}
\usepackage{graphicx}
\usepackage{enumitem}
\usepackage{gensymb}

\usepackage{multirow}%
\usepackage{amsmath,amssymb}%
\usepackage{amsthm}%
\usepackage{mathrsfs}%
\usepackage[title]{appendix}%
\usepackage{xcolor}%
\usepackage{textcomp}%
\usepackage{manyfoot}%
\usepackage{algpseudocode}%
\usepackage{listings}%

\usepackage{placeins}
\usepackage{float}
\usepackage{mathtools}

\DeclarePairedDelimiter{\nint}\lfloor\rceil

\makeatletter
\newcommand*{\textlabel}[2]{%
  \edef\@currentlabel{#1}
  \phantomsection
  #1\label{#2}
}
\makeatother

\usepackage[ruled,linesnumbered]{algorithm2e}
\makeatletter
\newcommand{\nosemic}{\renewcommand{\@endalgocfline}{\relax}}
\newcommand{\dosemic}{\renewcommand{\@endalgocfline}{\algocf@endline}}
\let\oldnl\nl
\newcommand{\nonl}{\renewcommand{\nl}{\let\nl\oldnl}}
\makeatother

\usepackage[style=numeric-comp, sorting=none]{biblatex}
\title{Magnetic control of magnetotactic bacteria swarms}

\author{
  Mihails Birjukovs\\
  MMML lab, Department of Physics\\
  University of Latvia (UL)\\
  Riga, Latvia, Jelgavas 3, 1004 \\
  \texttt{mihails.birjukovs@lu.lv} \\
  \And
  Klaas Bente\\
  Department of Biomaterials\\
  Max Planck Institute of Colloids and Interfaces\\
  Potsdam, Germany, Am M\"{u}hlenberg 1, 14476\\
  Institute for Molecular Medicine\\ 
  Health and Medical University Potsdam\\
  14471 Potsdam, Germany\\
  \And
  Damien Faivre\\
  Department of Biomaterials\\
  Max Planck Institute of Colloids and Interfaces\\
  Potsdam, Germany, Am M\"{u}hlenberg 1, 14476\\
  BIAM, Aix-Marseille University, CNRS, CEA\\ 
  Saint-Paul-lez-Durance, France, 13115\\
  \texttt{damien.faivre@cea.fr} \\
  \And
  Guntars Kitenbergs\\
  MMML lab, Department of Physics\\
  University of Latvia (UL)\\
  Riga, Latvia, Jelgavas 3, 1004 \\
  \texttt{guntars.kitenbergs@lu.lv} \\
  \And
  Andrejs Cebers\\
  MMML lab, Department of Physics\\
  University of Latvia (UL)\\
  Riga, Latvia, Jelgavas 3, 1004 \\
  \texttt{andrejs.cebers@lu.lv} \\
  }

\begin{document}
\maketitle

\begin{abstract}

Magnetotactic bacteria (MTB) are of significant fundamental and practical interest, especially for applications such as drug delivery and general-purpose object manipulators and payload carriers. While magnetic and other modes of control for individual MTB have been demonstrated, formation, motion and control of MTB swarms are much less studied and understood. Here, we present a torque dipole-based theoretical model for magnetic control of MTB swarms and two methods for swarm formation, and provide experimental validation of the proposed motion model. Model predictions are in good qualitative and quantitative agreement with experiments and literature. Additionally, we were able to determine the torque generated by \textit{Magnetospirillium gryphiswaldense} (MSR-1) MTB, and the value corresponds to the reported estimates reasonably well.

\end{abstract}

\keywords{Magnetotactic bacteria (MTB) \and swarm behavior \and magnetic control \and hydrodynamics with spin \and active matter}

\clearpage

\section{Introduction}\label{sec1}

Contemporary literature reports different interesting phenomena in chiral active matter, such as formation of rotating crystals of spinners \cite{cebers-ref-1,cebers-ref-2}, odd viscosity effects \cite{cebers-ref-2,cebers-ref-3}. Several approaches accounting for chirality in the continuum models of active matter are known. A quite general approach based on non-equilibrium thermodynamics of continuum media was developed in \cite{cebers-ref-4,cebers-ref-5}. There exist review papers on the subject -- for example, see \cite{cebers-ref-6,cebers-ref-7,cebers-ref-8}.

Magnetic micro-swimmers are of particular interest, with a natural directional self-propulsion mechanism and the potential to direct formation and emergent behaviors of swimmer ensembles, both controlled with applied magnetic field (MF) \cite{lauga-annurev-bacteria-hydrodynamics, mtb-review-paper-faivre, msr-1-switching-swimming-modes, mtb-rotating-field-theory-erglis-cebers, cebers-ref-11, ClusterEmergence, mtb-tunable-self-assembly-swarms, mtb-tunable-hydrodynamics, mtb-confined-convection, mtb-aps-colloquium-review-faivre-2024}. Among prospective applications, drug delivery and anti-cancer actuation stands out as especially attractive \cite{mtb-drug-delivery, mtb-drug-delivery-more-recent, mtb-cancer-treatment-hyperthermia}, and there is clear potential for more general-purpose object manipulation use cases \cite{ClusterEmergence,mtb-tunable-hydrodynamics, mtb-tunable-self-assembly-swarms, cebers-ref-11}. Motion of ensembles (swarms) of magnetotactic bacteria (MTB) due to flow induced by MTB torque dipoles is considered in this paper. A modification for a theoretical model developed in \cite{cebers-ref-11} is derived, and the updated model is validated against experimental data -- microscopically image motion of MTB swarms of different scales (sizes) driven by switched MF.

\section{A motion model for MTB swarms}\label{sec2}

Torques due to applied by MF in ensembles of magnetic particles causes various hydrodynamic phenomena \cite{mag-suspensions-eff-viscosity, ferrohydrodynamics-rosensweig, magnetic-fluids-cebers, cebers-prl-mag-drop-rotating-mf}, which are described in terms of hydrodynamics with spin -- the latter considers the antisymmetric stress determined by the number density of torques. Such models have been successfully applied to bacterial suspensions \cite{cebers-ref-1, cebers-hydrodynamics-with-spin}. Antisymmetric stress may appear when bacteria generate torque dipoles in the liquid -- one torque in a pair is due to a rotating flagellum, and the other, in the opposite direction, is caused by the MTB body rotation. In an orientationally ordered suspension, torque dipoles result in a \textit{couple stress} and induce flow of the suspension. This phenomenon arises in a suspension of MTB in an external field.

\begin{figure}[h]
\centering
\includegraphics[width=0.75\textwidth]{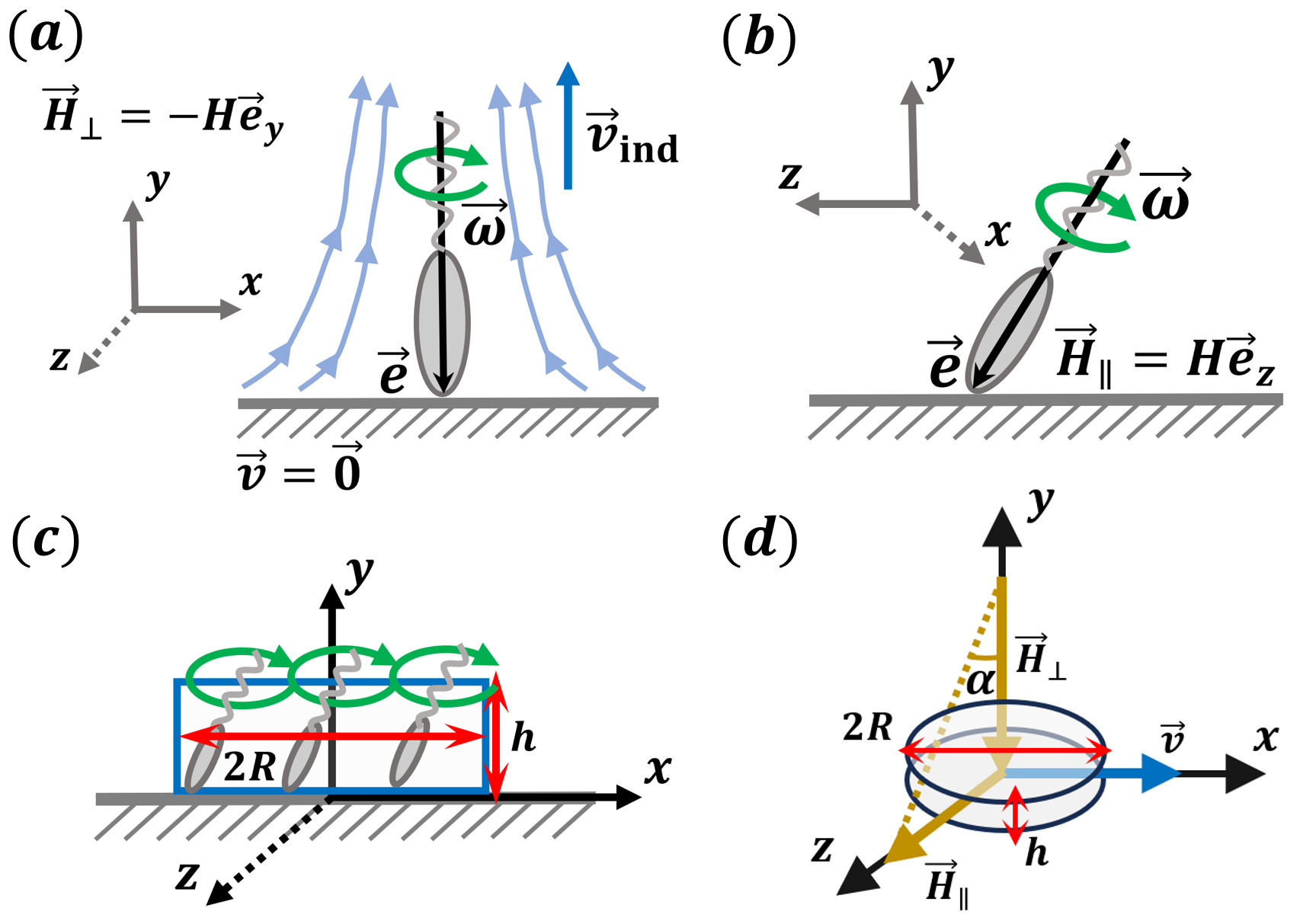} 
\caption{A schematic of the investigated system. (a) A single pusher-type MTB (its body and flagella represented with a gray ellipse and a gray wavy curve, respectively) with its self-propulsion direction $\vec{e}$ (note the coordinate axes) and the respective torque. If normal MF $H_\perp$ is applied, the MTB is propelled by its flagella motion (direction indicated by $\vec{\omega}$ and a green arrow) towards and stacked against a cell wall (no-slip boundary condition), where the propulsion force is compensated by the normal reaction of the wall, and the liquid surrounding the MTB is repulsed from the wall (light-blue lines show the liquid flow streamlines about the MTB, $\vec{v}_\text{ind}$ indicates flow general direction). (b) MTB inclination due to torque along $x$, induced by applied surface-parallel MF ($H_\parallel$). (c) Torque acting on the MTB layer (swarm) boundary; the (idealized) cylindrical swarm has a diameter $2R$ and a height $h$. (d) MTB swarm motion direction $\vec{v}$ due to applied $H_\parallel$ (with $H_\perp$ keeping the swarm intact), given by the left-hand rule. Here $\alpha$ is the angle between $\vec{H}_\perp$ and $\vec{H}_\perp + \vec{H}_\parallel$.}
\label{fig:theory-swarm-motion-sketch}
\end{figure}

Previously, a linear model for MTB swarm motion in MF was established for MSR-1 (\textit{Magnetospirillium gryphiswaldense} \cite{mtb-msr1-og-paper-90s}), relating swarm velocity to applied MF \cite{cebers-ref-11}. Swarms were defined as clusters of MTB, stable under applied static MF $H_\perp$ orthogonal to the surface(-s) of the MTB suspension cell where MTB are deposited (Figure 1 in \cite{cebers-ref-11} and Figures \ref{fig:theory-swarm-motion-sketch}c-d). While the model correctly predicted the MTB swarm motion directions, more recently published experimental data regarding MSR-1 \cite{swimming-organism-data-bank} and newly developed image analysis methods (Appendix \ref{secA3}) indicated that we have overestimated swarm thickness to radius ratio (a parameter of the model). In addition, we have now come up with a simpler model relating MTB swarm velocity magnitude to the MTB-induced flow field about the swarm, accounting for how the swarm interacts with the surface along which it travels under the effect of applied MF.

Couple stress induced in a bacteria suspension due to the action of torque dipoles can be derived in the continuum mechanics approximation (coarsening of an ensemble of entities to a distribution of torque dipoles). A pusher bacterium with a direction of self-propulsion $\vec{e}$ applies torques to the liquid $- \tau\vec{e}$ and $\tau\vec{e}$, acting from the tail (flagellum) and the head (body) of the bacterium, respectively \cite{cebers-ref-9} (Figure \ref{fig:theory-swarm-motion-sketch}a). The torque acting on the volume of suspension enclosed by a surface $S$ is created by bacteria with $\vec{e}$ crossing the surface elements $\Delta S\vec{n}$. Then the torque $\delta\vec{T}$ acting due to bacteria crossing the surface element $\Delta S\vec{n}$ equals $\delta\vec{T}=-\tau\vec{e} \cdot bn(\vec{e}\cdot\vec{n})\Delta S $, where $b$ is the distance between the points where torques $-\tau\vec{e}$ and $\tau\vec{e}$ are applied (i.e., proportional to the MTB body length), and $n$ is the volumetric bacteria number density. For the couple stress $j_{ik}$, with $\delta T_{i}=j_{ik}n_{k}\Delta S$, this yields $j_{ik}=-\tau bne_{i}e_{k}$. Due to the angular momentum conservation for the torque dual to the antisymmetric part of the stress tensor, one has $\sigma_{i}=e_{ikl}\sigma_{kl}=\partial j_{ik}/\partial x_{k}$ (internal spin angular momentum is neglected) \cite{cebers-ref-4,cebers-ref-5}. This relation shows that when torque dipoles are at an angle to a free boundary, they induce flow \cite{cebers-ref-10, cebers-ref-11}. This mechanism was applied to describe the observed motion of MTB swarms in oblique MF in \cite{cebers-ref-11,cebers-ref-12}.

Let us illustrate this mechanism in the 2D case, which allows for an analytical solution \cite{cebers-ref-12}. In MF perpendicular to the boundaries of the cell ($\vec{H}_\perp$), bacteria are stacked against the wall ($\vec{e}=-\vec{e}_{y}$), as shown in Figure \ref{fig:theory-swarm-motion-sketch}a. If the tangential field $\vec{H}_\parallel$ is applied along the $z$ axis, then the torque acting on the liquid from the tail of the bacterium has a component along the $x$ axis (Figure \ref{fig:theory-swarm-motion-sketch}b). The couple stress acting on the layer of suspension with a thickness $h$ then reads ($e_{z}>0;e_{y}<0$):

\begin{equation}
j_{zy}=-\tau bne_{z}e_{y}\theta(h-y)\theta(y)
\label{eq:couple-stress}
\end{equation}
where $\theta(x)$ is the Heaviside function.

Thus, the torque acting on the layer boundary is $\sigma_{z}=\partial j_{zy}/ \partial y=\tau bn e_{z}e_{y}\delta(y-h)$ (Figure \ref{fig:theory-swarm-motion-sketch}c). The $\delta$ function singularity is compensated by the corresponding singularity of the viscous stress and, as a result, the velocity field has a discontinuity on the surface of suspension $v_{x}(h+0)-v_{x}(h-0)=-\tau bn e_{z}e_{y}/2\eta$. Since $\sigma_{z}<0$, liquid at $z>h$ moves in the $x>0$ direction. This direction of liquid and swarm motion is according to the left-hand rule (LHR), as seen in Figure \ref{fig:theory-swarm-motion-sketch}d -- if $\vec{H}_{\perp}$ enters the palm, and four fingers are in the direction of $\vec{H}_{\parallel}$, then the thumb indicates the direction of swarm motion, in this case along the $x>0$ direction. If the direction of $\vec{H}_{\parallel}$ is reversed, so is the direction of swarm motion \cite{cebers-ref-11}.

Interestingly, LHR could also potentially be applied in the case of vortex flow in droplets containing MTB \cite{cebers-ref-13}. According to \cite{cebers-ref-13}, North-seeking (moves along MF) MTB accumulate on the surface of a droplet and clockwise circulation arises. This corresponds to the LHR -- if MF enters the palm, the fingers indicate the gravity force direction, then the thumb points to the direction of induced flow, which explains the observed clockwise circulation. In this case, the oblique orientation of the torque dipole is caused by the gravity force.

A quantitative description of the flow induced by torque dipoles may be obtained, considering the swarm in the form of an infinite stripe (thickness $h$ and width $2R$) on top of the wall with a non-slip boundary condition \cite{cebers-ref-11}. Bacteria remain bound in the swarm due to the action of normal MF \cite{mtb-tunable-self-assembly-swarms}, and it is assumed that MF is strong enough to orient all MTB. The boundary conditions for the Stokes equation are non-slip boundary condition on the wall, the jump of the velocity at $y=h;-R<x<R$ with the corresponding condition far from the swarm. In this case, the solution of 2D problem may be obtained using methods of functions of complex variable (for example, see \cite{cebers-ref-14}). The solution of the Stokes equation in the $\text{Im}(z)>0$ half plane is expressed by two analytical functions $\chi,\varphi$

\begin{equation}
v_{x}-iv_{y}=i\left(\chi'+\bar{\varphi}+\bar{z}\varphi'\right)
\label{Eq:2}
\end{equation}
The solution satisfying the non-slip boundary conditions at $y=0$ and velocity jump at $y=h;~-R<x<R$
reads

\begin{equation}
v_{x}-iv_{y}=\frac{\Delta v}{2\pi i} \left(\int^{R}_{-R}\frac{dx'}{ih+x'-z}-\int^{R}_{-R}\frac{dx'}{ih+x'-\bar{z}}\right)+2y\varphi'
\label{Eq:3}
\end{equation}
where

\begin{equation}
\varphi'=\frac{\Delta v}{2\pi}\int^{R}_{-R}\frac{dx'}
{(x'-ih-z)^2}
\label{Eq:4}
\end{equation}

Flow streamlines according to the relations $v_x(x,y,R)$ and $v_y(x,y,R)$, given in Appendix \ref{secA1}, are shown in Figure \ref{fig:theory-flow-field-norm-velocity}a. Swarm motion is characterized by 

\begin{equation}
v_s(h,R) = 1/2\cdot \left( v_{x}(0,h+0,R)+v_{x}(0,h-0,R) \right)
\label{eq:velocity-correction}
\end{equation}
Its normalized value $v_s(1,R/h)$ is shown in Figure \ref{fig:theory-flow-field-norm-velocity}b. With this and the previously established LHR, the swarm velocity can be expressed as

\begin{equation}
\vec{v} = v_s (h,R) \cdot v_0 \left( \vec{e}_{H_\parallel} \times \vec{e}_{H_\perp} \right)
\label{eq:velocity-expression-final}
\end{equation}
where $v_s$ is given by (\ref{eq:velocity-correction}), and $v_0$ is as derived in \cite{cebers-ref-11}:

\begin{equation}
v_0 = \frac{b \tau}{8 \eta} \cdot n \sin{(2 \alpha)}
\label{velocity-baseline}
\end{equation}
where $b$ is proportional to the MTB body length, $\eta$ is the dynamic viscosity of the medium, and $\alpha$ is the angle relating $\vec{H}_\perp$ and $\vec{H}_\perp + \vec{H}_\parallel$, as shown in Figure \ref{fig:theory-swarm-motion-sketch}d.

\begin{figure}[H]
\centering
\includegraphics[width=0.90\textwidth]{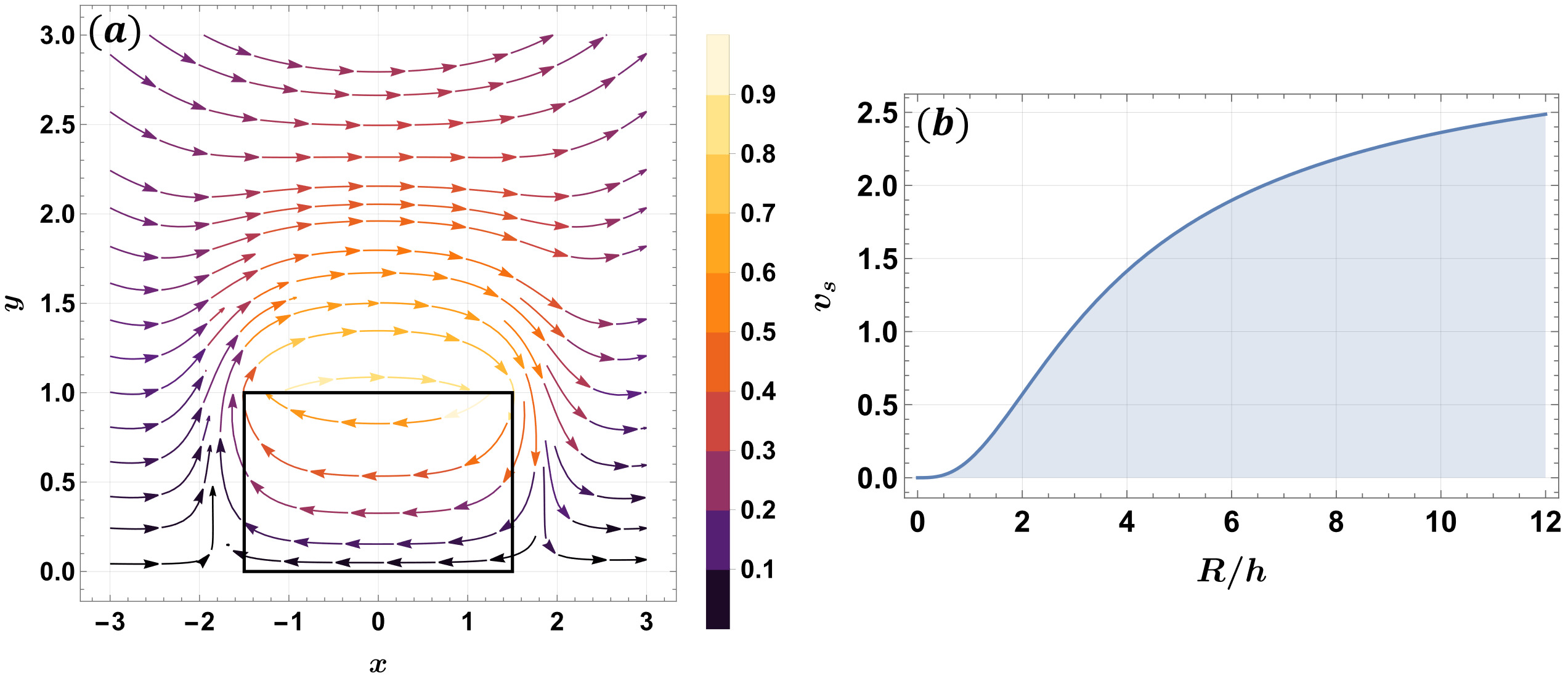} 
\caption{(a) Velocity field streamlines (axes as in Figure \ref{fig:theory-swarm-motion-sketch}c.) induced by torque dipoles -- velocity components are normalized by $\Delta v/2\pi$, with $R/h=1.5$ as an example case (the black rectangle is the swarm cross-section). (b) Normalized velocity of the swarm $v_s$ versus the relative swarm width $R/h$.}
\label{fig:theory-flow-field-norm-velocity}
\end{figure}

\section{Experimental validation}\label{sec3}

\subsection{Swarm formation}

\begin{figure}[htbp]
\centering
\includegraphics[width=0.75\textwidth]{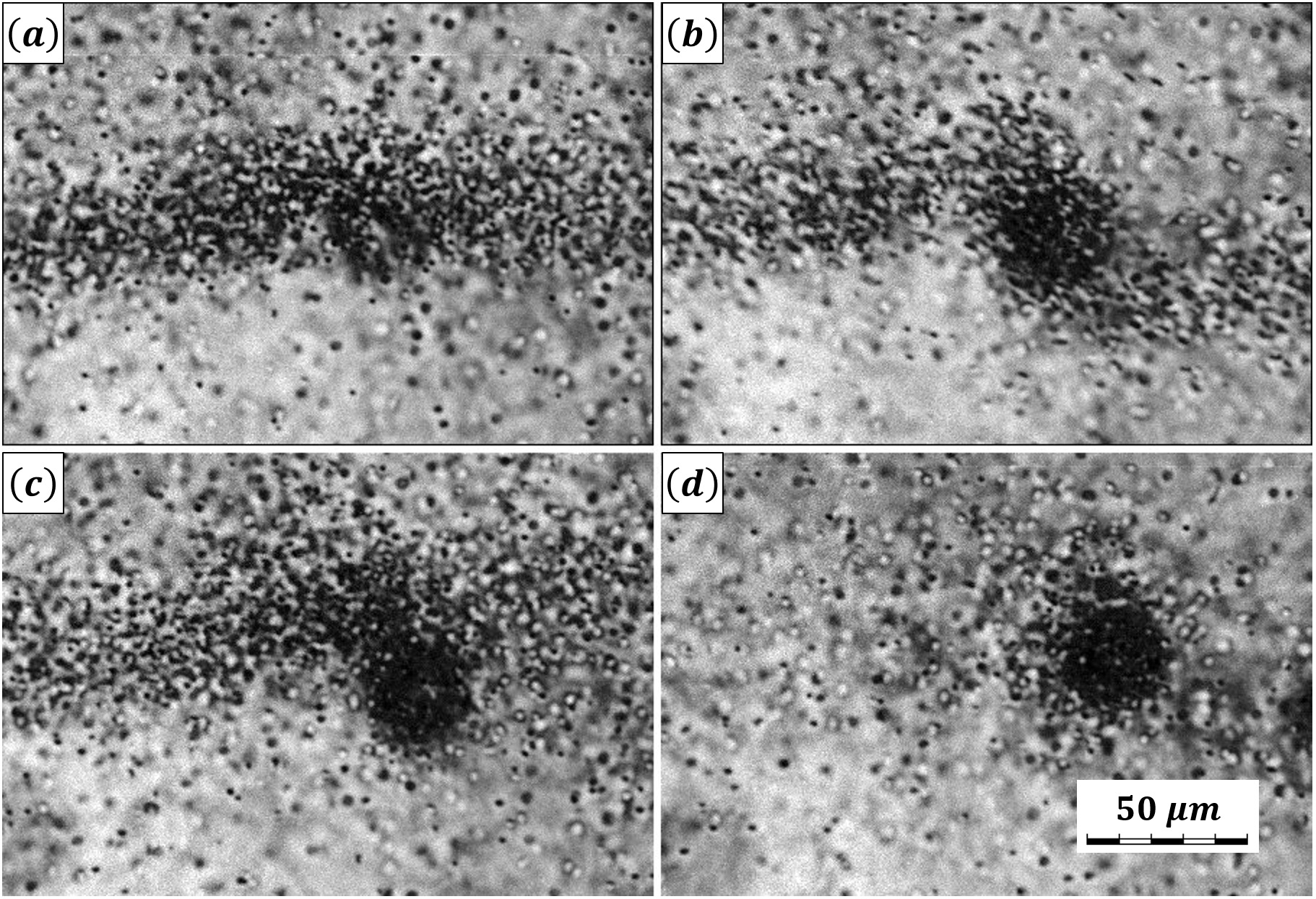} 
\caption{Swarm formation from a band of MTB: (a) the MTB band is forced towards the wall with $H_\perp=-30~Oe$; (b) enabled $f=1~Hz$ rotating MF $H_\text{rot}=30~Oe$ in the $XY$ (image) plane, which distorts the band; (c) the swarm is formed and remains stable after $H_\text{rot}$ is disabled, with a part of the MTB band still near the swarm; (d) the MTB band is removed by applying $\lambda=470~nm$ (blue) light.}
\label{fig:swarm-formation-1}
\end{figure}

\begin{figure}[htbp]
\centering
\includegraphics[width=1.0\textwidth]{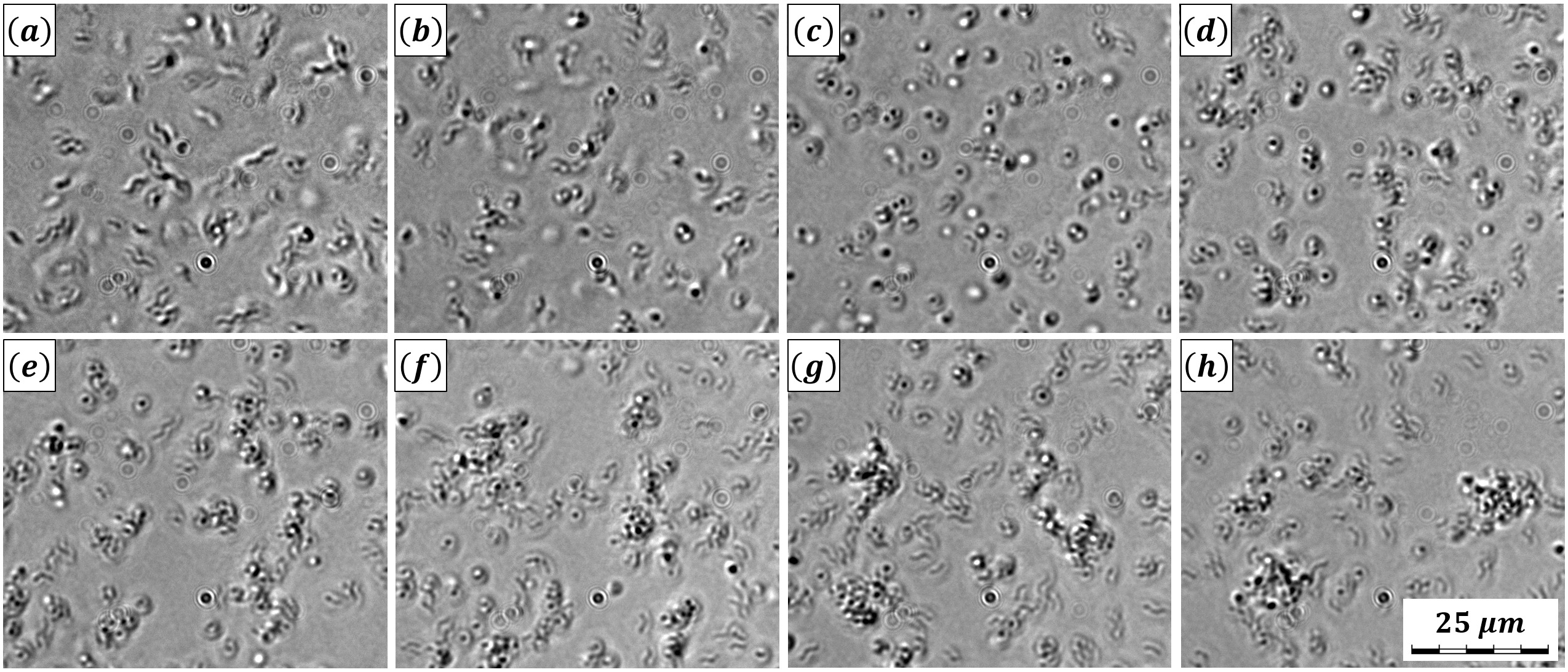} 
\caption{Spontaneous swarm formation. At $t=1.6~s$ (b), $H_\perp=-50~Oe$ MF is switched on. Timestamps: (a) $t=0~s$, (b) $t=1.6~s$, (c) $t=1.7~s$, (d) $t=2.8~s$, (e) $t=3.35~s$, (f) $t=4.65~s$, (g) $t=6.55~s$, (h) $t=9.4~s$. Note the two stable larger swarms present in (h) formed within roughly $7.8~s$.}
\label{fig:swarm-formation-2}
\end{figure}

To validate the updated theoretical model, a series of experiments were performed with MTB swarms of different scales/sizes. A microscopy setup with MF coils like the one described in \cite{microscopy-setup-faivre} was used, and samples were prepared as described in \cite{cebers-ref-11} (briefly summarized here in Appendix \ref{secA2}). There are two feasible methods to create swarms in prepared suspensions, illustrated in Figure \ref{fig:swarm-formation-1} and Figure \ref{fig:swarm-formation-2}.

The first method exploits large bacteria concentrations in MTB bands formed in the suspension \cite{cebers-ref-11, mtb-band-formation-erglis-cebers}. When a band is exposed to perpendicular MF ($H_\perp=-30~Oe$), it splits into two parts, according to the number of north- and south-seeking MTB in the sample population \cite{FaivreBiophysJ, mtb-confined-convection}, and forms dense bands close to the surfaces of the MTB suspension cell (Figure \ref{fig:swarm-formation-1}a). Applying additional rotating MF ($H_\parallel=30~Oe$, $f=1~Hz$) distorts the band and forms a swarm, which continues to exist after the rotating MF is turned off (Figures \ref{fig:swarm-formation-1}b-c). However, this manipulation assembles only a part of the band MTB into the swarm, and the band remains -- this could be due to bacteria power output decay, observed in \cite{BlueLight}. Introducing blue light ($\lambda=470~nm$, \textit{CoolLED Ltd.}, pE-100) is sufficient to move the band away from the swarm (effects of illumination on bacteria are explored in-depth in \cite{bacteria-polygonal-spinning-motion-light-induced}). As a result, the swarms are separated from the band and can be further manipulated, as seen in Figure \ref{fig:swarm-formation-1}d. This is how we initialized larger-scale swarms.

However, there is another, simpler method. Only two things are required -- a high enough MTB concentration and perpendicular MF. Some time after MF is enabled, MTB spontaneously form swarms, as shown in Figure \ref{fig:swarm-formation-2}, using smaller-scale swarms as an example. Here, MTB did not form a band. After $H_\perp=-50~Oe$ was applied (Figure \ref{fig:swarm-formation-2}b), MTB were first oriented normally to the suspension cell surfaces (b-c), and then swarms started to appear at the surfaces. Initially, the swarms do not have a clear form, and MTB chains/networks are observed (e-g), but over time they become roughly circular (h). A very similar observation was recently made by Pierce \textit{et al.} \cite{ClusterEmergence, mtb-tunable-self-assembly-swarms} -- it was shown that two MTB, when forced to orient normally to a wall, experience an attractive hydrodynamic interaction, and this leads to MTB clustering into (eventually circular) ensembles with greater concentration, as clearly seen in snapshots (c-h) from Figure \ref{fig:swarm-formation-2}.

\begin{figure}[htbp]
\centering
\includegraphics[width=0.80\textwidth]{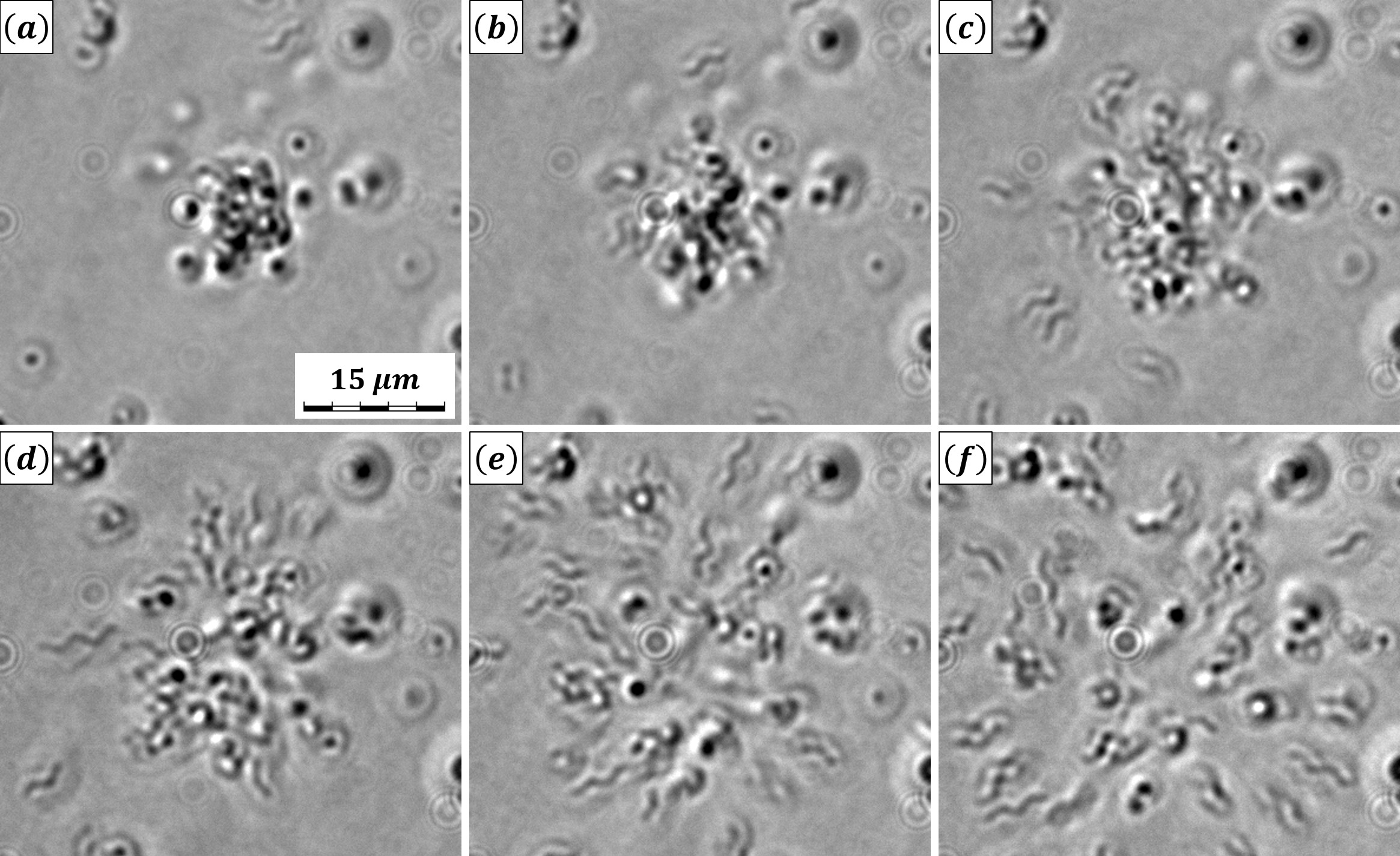} 
\caption{Swarm breakup: (a) an intact swarm right before $H_\perp$ is switched off and (b-f) $150~ms$, $250~ms$, $350~ms$, $450~ms$ and $550~ms$ after $H_\perp$ is disabled.}
\label{fig:swarm-disruption}
\end{figure}

Figure \ref{fig:swarm-disruption} further illustrates that plane-orthogonal MF determines swarm stability. A smaller-scale swarm was formed and manipulated with plane-parallel MF before all MF components were switched off. Figure \ref{fig:swarm-disruption}a shows a snapshot of a swarm just before MF is disabled -- observe in (b-f) how the swarm rapidly breaks up, and MTB are reoriented, within less than a second.

\subsection{Swarm motion control}

With clear and reliable methods for swarm formation and MTB suspension preparation, one can assemble swarms of various sizes and manipulate them with MF. To validate the theory developed in \cite{cebers-ref-11} and here in Section \ref{sec2}, swarms of two different scales (i.e., expected mean sizes) are observed in a series of experiments (i.e., image series acquisitions) where MF is switched to different configurations to verify both the swarm/MF direction correspondence, and the theoretical predictions for swarm velocity magnitude.

The methods used/developed for image processing and data analysis 
are briefly described in Appendices \ref{secA3}, \ref{secA4} and \ref{secA5}. For every image sequence recorded, swarms are detected and tracked over time, and their velocity time series are recovered. Angle $\alpha$ time series for MF are known, and MTB concentration in swarms is estimated from its distribution about the swarms, assuming swarms are single layers of MTB (Figure \ref{fig:theory-swarm-motion-sketch}).

\begin{figure}[htbp]
\centering
\includegraphics[width=1\textwidth]{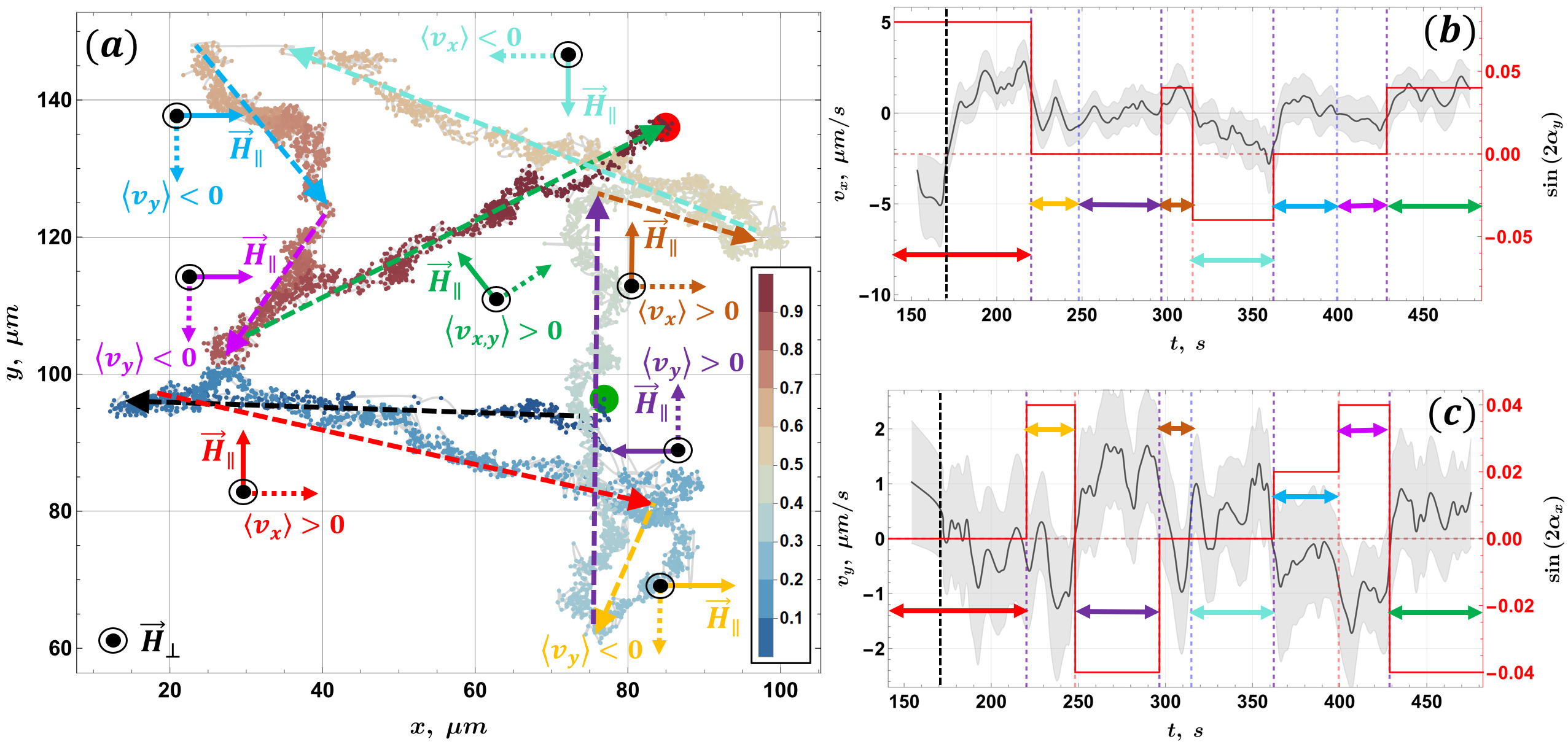} 
\caption{MTB swarm motion in applied MF $\vec{H} = \vec{H}_\perp + \vec{H}_\parallel$: (a) a swarm trajectory over $10 \text{K}$ images with highlighted positions at every time stamp, color coded by order of appearance (legend in the top-right corner), with constant $\vec{H}_\perp$ directed towards the viewer of the $XY$ plane (indicated in the bottom left corner); (b-c) swarm horizontal ($x$) and vertical ($y$) velocity components (black curves with gray uncertainty bands, Appendix \ref{secA5}) versus applied MF (solid red lines), expressed in terms of $\alpha_k$, $k = \{x,y\}$ (please see Figure \ref{fig:theory-swarm-motion-sketch} for angle $\alpha$ definition). Dashed colored lines in (a) correspond to constant MF intervals indicated in (b-c) with respective colors. MF switch time stamps are marked in (b-c) with dashed vertical lines: red for the $\vec{H}_\parallel$ component that affects the respective $\vec{v}$ component, blue for the one that does not, and purple for when both $\vec{H}_\parallel$ components are switched. Note the abrupt swarm turns driven by MF switching. For each trajectory interval highlighted in (a), corresponding $\vec{H}_\parallel$ directions (solid arrows), expected $\vec{v}$ (motion forcing) directions (dashed arrows), and mean experimental velocity signs are annotated. Trajectory start/end points are marked with a green/red circles, respectively. Note the dashed black arrow in (a) and dashed vertical black lines in (b,c) -- these indicate the trajectory and time series intervals, respectively, where the considered swarm interacted with and absorbed another, smaller one.}
\label{fig:experiment-swarm-motion-direction-2}
\end{figure}

An example swarm trajectory is shown in Figure \ref{fig:experiment-swarm-motion-direction-2}a. Here, an image sequence of $10 \text{K}$ frames was analyzed and a trajectory of a single small-scale swarm was reconstructed from detection events in each frame where it appeared -- positions at every instance are highlighted with dots, color coded in order of appearance (color bar in Figure \ref{fig:experiment-swarm-motion-direction-2}a). Figures \ref{fig:experiment-swarm-motion-direction-2}b and \ref{fig:experiment-swarm-motion-direction-2}c show the velocity component time series and the dynamics of the MF components that drive respective velocity components. Constant MF intervals in (b,c) are indicated with horizontal arrows, and arrows of corresponding colors in (a) outline the mean directions of swarm motion for the MF intervals. Note that the experimentally observed swarm motion direction matches theoretical predictions very well, especially for $t \in [170;220]~s$, $t \in [248;296]~s$ and $t \in [429;482]~s$ (b,c), and, even in cases where drift in directions not enforced by applied MF, the velocity component predicted to be dominant due to MF is the one largely determining the swarm trajectory.

Clearly, though, the observed drift is significant enough to warrant discussion. We find that it cannot be attributed to swarm inertia for any of the cases, especially, for smaller scale swarms (Figure \ref{fig:experiment-swarm-motion-direction-2} is one such example) -- not only do we consistently see virtually instantaneous motion direction change as MF is switched, but also the excess drift directions are not compatible with this explanation. However, there are several more plausible reasons for this swarm behavior.

One potential and simple explanation is that, aside from the tracked swarms that are above a predefined size threshold (constant across experiments at similar swarm scales), there are always smaller formations (clusters/chains) of MTB, spontaneously forming and disbanding within the ambient MTB concentration field through which the swarms travel. Proximity to such small MTB formations can affect swarm trajectories, as the latter hydrodynamically interact with the former, inducing mutual attraction and thus deviation from the motion enforced by applied MF. This is different from cases where two significant swarms interact -- as seen in Figure \ref{fig:experiment-swarm-motion-direction-2} (regions of interest indicated with black dashed lines), this results in blatant violations of the LHR (initial $\langle v_x \rangle < 0$, as opposed to the expected $\langle v_x \rangle > 0$), whereas interactions with small MTB formations simply introduce marginal drift. However, the observed drift is too consistent over time, and the above can only account for lower-wavelength perturbations in trajectories.

Interestingly, in our case, greater $H_\parallel$ magnitude corresponds to less significant trajectory drift with respect to the magnetically forced motion direction. This is especially evident for $t \in [429;482]~s$ (Figures \ref{fig:experiment-swarm-motion-direction-2}b,c), where $H_\parallel$ is maximal over the trajectory shown in Figure \ref{fig:experiment-swarm-motion-direction-2}, and the swarm almost perfectly follows the theoretically expected path, with very little dispersion about the mean roughly rectilinear path. Notably, drift in directions not prescribed by MF was also reported by Pierce \textit{et al.} in \cite{mtb-tunable-self-assembly-swarms}, specifically under weaker MF, where MTB clusters translated at an angle to the motion direction enforced by MF. However, Pierce \textit{et al.} also report that the extent of deflection also depends on the tilt angle for MF \cite{mtb-tunable-hydrodynamics} ($\alpha$), and thus the ratio of $H_\parallel$ and $H_\perp$ magnitudes -- this was attributed to the lateral movement of tilted MTB near the cell surface due to competing hydrodynamic and magnetic torques acting on the MTB body \cite{mtb-tunable-self-assembly-swarms, ClusterEmergence, mtb-tunable-hydrodynamics} (the latter reference provides the more detailed explanation of this phenomenon).

Another factor to consider is swarm shape and size stability. For smaller swarms the shape, while roughly circular on average, is generally less stable, and the swarm is less inertial, more prone to being affected by ambient MTB concentration perturbations. With larger swarms, one often observes that their shapes can become slightly elliptic as they travel, elongated in the direction of motion, with trails of MTB ejected from the swarm and left behind, and new MTB along swarm paths constantly integrated into the swarms. This elongation is in line with observations made in \cite{mtb-tunable-self-assembly-swarms}, where at MF of $\sim 50~ Oe$ (corresponds to $H_\perp$ applied for large swarms in our case) MTB clusters were distorted in the direction of motion, explained through MTB dispersion due to a wide range of individual MTB velocities and instantaneous swimming directions.

It is interesting to note, however, that for larger swarms some of the observed dispersion could potentially be due to MTB updraft near the center of the swarm -- we observe an average radial MTB influx at the swarm boundaries, with radial outward flow of MTB in a plane above the swarm, which is something that was directly observed and quantified in \cite{mtb-confined-convection}. Average radial net influx of MTB toward the swarm is also consistent with the models for swarm formation (Figure \ref{fig:theory-swarm-motion-sketch}) \cite{ClusterEmergence, mtb-tunable-self-assembly-swarms}. Some of the MTB ejected upwards may end up in the MTB trails left behind the swarms due to downward (with respect to the surface along with swarms travel) flow about the updraft current (visualized in \cite{mtb-confined-convection}), and could falsely appear and be interpreted as though they were originally ejected from the swarm laterally. This is, however, only a concern for the larger swarms -- we did not observe updraft for smaller-scale swarms.

Aside from swarm shape perturbations, another factor potentially promoting trajectory point dispersion about mean motion directions for constant MF intervals (Figure \ref{fig:experiment-swarm-motion-direction-2}a) is partial MTB orientation/motion incoherence. According to \cite{ClusterEmergence}, our swarms, while expected to be stable, are rather close to the order/disorder phase boundary of the MTB swarm number density and MF strength phase diagram proposed therein, meaning our swarms likely exhibit a wide range of instantaneous MTB orientations and swimming directions. In addition, one must recall that MTB magnetic moment, unlike theoretical assumptions, is can realistically be misaligned with respect to the MTB body orientation. If magnetic moment misalignment statistics for MSR-1 (no such data currently available, to our knowledge) are anything like \textit{Magnetospirillum magneticum} (AMB-1 \cite{mtb-amb1-og-paper-90s}), for which, very roughly speaking, MTB axis and magnetic moment misalignment angle is $\sim 6.5^{\circ}$ with a standard deviation of $\sim 3.2^{\circ}$ \cite{mtb-magnetic-moment-misalignment}, they could have a tangible effect on swarm stability and motion control. Moreover, magnetosome chains in MSR-1 are known to exhibit magnetosome magnetic moment misalignment with respect to the chain axis \cite{mtb-msr1-magnetosome-moment-misalignment-in-chains}. Depending on average magnetic moment and MTB body axis misalignment for MSR-1, one could potentially observe both trajectory oscillations and significant swarm drift in directions not enforced by MF. One should also keep in mind that individual MTB parameters (length, torque, etc.) may vary within a population. Finally, while very unlikely a concern in our experiments, one must always keep in mind that light intensity/non-uniformity at the observed sample could be an issue and interfere with MTB motion, as demonstrated in \cite{bacteria-polygonal-spinning-motion-light-induced}.

Having verified swarm motion directions correspond to MF components as expected due to theory, with reasonable/explicable deviations, one should also check if the velocity magnitudes scale linearly with MF magnitude. MTB body length is set to $b = 3.85 ~\mu m$ according to \cite{swimming-organism-data-bank}, which matches the value measured from the acquired images (Figure \ref{fig:swarm-disruption}f is a good example where full MTB lengths are observed).
Assuming that all swarms are single-MTB layers, one has $h=b$.
Thus, $n$ can be estimated from surface (i.e., image plane) number density $n_s$ as $n = n_s/h$, where $n_s$ is determined from images based on the radial $n_s$ distribution of MTB around the cores, since, in this case, direct measurements in densely packed cores is generally not feasible with our microscopy setup. However, we argue that under the single MTB layer swarm assumption, this approach is reasonable. Moreover, \textit{a posteriori}, with $n_s \in ~\sim (0.015;0.042)~{\mu m}^{-2}$ and given $H_\perp \in [30;50]~Oe$, according to the $(n_s,H)$ phase diagram obtained experimentally for AMB-1 (arguably morphologically similar to MSR-1 \cite{mtb-magnetic-moment-misalignment}) MTB swarms/clusters in \cite{ClusterEmergence}, our experiments are within the threshold where this $n_s$ measurement method should yield number density close to that within swarm cores, a not a lower bound that significantly underestimates $n_s$. Both larger and smaller swarms exhibit ordered (i.e., clustered) motion and structure, which is in line with the prediction due to the phase diagram in \cite{ClusterEmergence} for our $n_s$ and $H_\perp$.

\begin{figure}[htbp]
\centering
\includegraphics[width=0.90\textwidth]{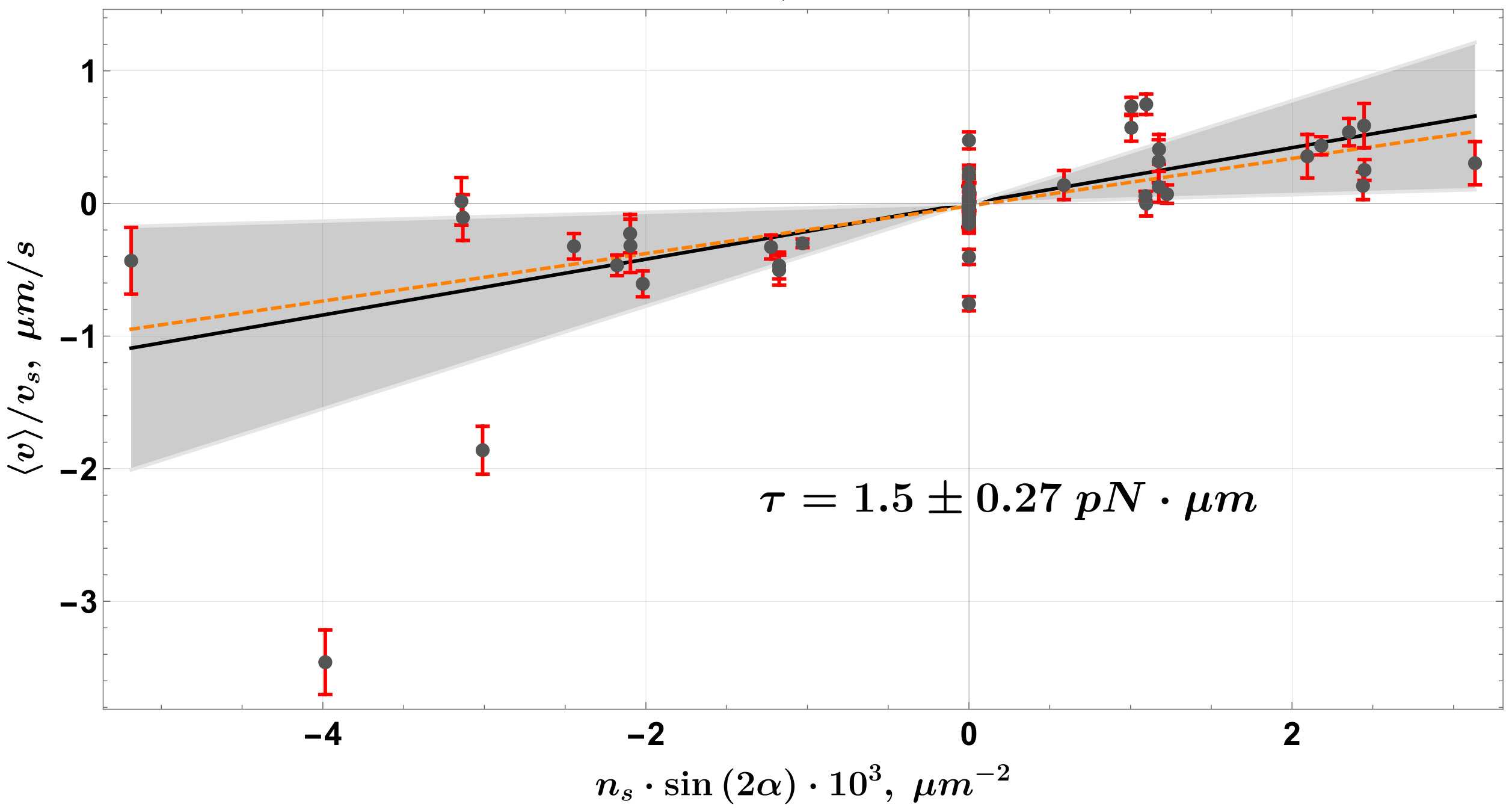} 
\caption{Aggregated $\langle v \rangle$ component values (gray points with red error fences) adjusted by $v_s$ (\ref{eq:velocity-correction}) for both velocity components in image coordinates versus their respective driving $\vec{H}_\parallel$ components (Figures \ref{fig:experiment-swarm-motion-direction-2}b-c) expressed as $\alpha_k$, $k = \{x,y\}$ (Figure \ref{fig:theory-swarm-motion-sketch}), and the estimated swarm MTB surface number density $n_s$. The black solid line is the linear fit for the data with light gray slope error bands, and the dashed orange line is a $0.5$ quantile linear fit, for reference. Note the estimated MTB dipole torque $\tau$.}
\label{fig:experiment-theory-fit}
\end{figure}

With this, one can fit (\ref{eq:velocity-expression-final}) to experimental data, represented by mean velocity $\langle v \rangle$ for each constant $\vec{H}$ interval (Figure \ref{fig:experiment-swarm-motion-direction-2}b-c) for every observed swarm, and determine the MTB dipole torque $\tau$ in the process:

\begin{equation}
\langle v \rangle = v_s \cdot v_0;~~ v_0 = \frac{b \tau}{8 \eta} \cdot \frac{n_s}{h} \cdot \sin{(2 \alpha)} \Rightarrow
\frac{\langle v \rangle}{v_s} = C_1 \cdot n_s \sin{(2 \alpha)}
\Rightarrow \tau = 8 \eta C_1
\label{eq:velocity-fit-theory-experiment}
\end{equation}
where $C_1$ is the linear fit constant (slope) and $\eta = 8.9 \cdot 10^{-4}~Pa\cdot s$. Figure \ref{fig:experiment-theory-fit} represents $\langle v \rangle$ pooled from all experiments with a linear fit according to (\ref{eq:velocity-fit-theory-experiment}), where $v_0 (\alpha): ~ v_0 (0) = 0$ (no directed motion with $H_\parallel = 0$).

The linear fit slope error is $\delta C_1 = 17.7 \%$, with fit quality measured by $R^2 = 0.37$, which is due to the two outlying points in the 3-rd quadrant of Figure \ref{fig:experiment-theory-fit}. These two points correspond to one of the experiments involving a large swarm driven by MF. The most likely issue is that, while a single MTB layer approximation is very much valid for small swarms, this particular large swarm apparently has thickness $h$ significantly in excess of MTB length $b$. Adjusting $h$ to be some multiple $>1$ of $b$, we observe the two outliers become part of the main data point cluster, dramatically improving $R^2$ and marginally reducing $\delta C_1$ (the relatively small impact of these outliers on the $C_1$ is indicated by the $0.5$ quantile linear fit, which is very close to the regular linear fit). However, since we have no means of reasonably estimating $h$ outside the single layer assumption, we leave the results as they are, for consistency. Note also a vertically spread cluster of data points at $\alpha = 0$ -- while many are near-zero $\langle v \rangle$ instances, there are points with significant deviations of velocity components from zero, where MTB drift was observed in directions not imposed by applied MF. The reasons for drift with applied $\vec{H}_\parallel$ MF were already discussed above, but without $\vec{H}_\parallel$ MF MTB can still drift quasi randomly quite significantly along the cell surface \cite{mtb-tunable-hydrodynamics}, and the errors in $\langle v \rangle / v_s$ are due to oscillatory motions of MTB swarms -- possible causes were covered above as well.

With this, we estimate the MTB dipole torque $\tau = 1.5 \pm 0.27~ pN \cdot \mu m$. To our knowledge, there currently isn't a known $\tau$ value for MSR-1 (there are \textit{thrust} measurements for related AMB-1 MTB, but not torque \cite{BlueLight, mtb-aps-colloquium-review-faivre-2024}), so it makes sense to at least check for any potentially relevant estimates. Some are given in \cite{lauga-annurev-bacteria-hydrodynamics}: stall torque measurements using optical tweezers \cite{bacteria-torque-measurements-optical-tweezers-stall} yielded $\tau = 4.6~ pN \cdot \mu m$ (uncertainty unspecified); using beads attached to flagella to measure torque in motion \cite{bacteria-torque-measurements-beads-motion} resulted in $\tau = 1.26 \pm 0.19~ pN \cdot \mu m$; theoretical estimates in \cite{bacteria-torque-theoretical-estimate} produced an estimate about an order of magnitude lower than the two above instances, $\tau = 0.37 \pm 0.1~ pN \cdot \mu m$. Clearly, our result is within the right order of magnitude with respect to these estimates, and the closest value to our own is $\tau = 1.26 \pm 0.19~ pN \cdot \mu m$ versus the new $\tau = 1.5 \pm 0.27~ pN \cdot \mu m$, which have significantly overlapping uncertainty intervals. However, we do not assume this makes either result definitive, or the other two previous estimates off. Our result is of the indirect kind, specifically for MSR-1. Clearly, it would be of interest to conduct more torque measurements, direct and indirect.

\section{Conclusions \& outlook}\label{sec13}

To summarize, we have presented a theoretical model for MTB swarm motion along a surface in oblique MF, and validated it experimentally. We have shown that MTB swarms move according to the left-hand rule (Figure \ref{fig:theory-swarm-motion-sketch}d). Observed swarm motion is in qualitative and quantitative agreement with the linear model (swarm velocity magnitude proportional to the motion plane-parallel MF magnitude, Figure \ref{fig:theory-swarm-motion-sketch}), with the couple stress derived by coarsening an ensemble of entities to a distribution of torque dipoles (a continuum mechanics approximation). We were also able to derive the torque generated by the MRS-1 MTB rotary motors, and it is in reasonable agreement with the relevant data available in literature. We find that the oscillations in MTB swarm trajectories, as well as swarm drift in directions not directly enforced by applied MF, are within expectations and agree with observations made by other research groups.

We have also reported two methods for MTB swarm formation, of potential use for on-demand and controllable swarm production for subsequent control. However, aside from \cite{ClusterEmergence, mtb-tunable-self-assembly-swarms}, there have been very few studies exploring MTB swarm formation systematically, and there is yet much work to be done. That is, to approach general-purpose applications of MTB as payload carriers or volumetric mixers (exploiting three-dimensional flow structures about MTB swarms \cite{mtb-confined-convection}), there must be a clear understanding of how MF configuration and switching sequences, as well as MTB concentration and other sample/environment parameters, affect size and stability of newly formed swarms. It is further of interest to explore inter-swarm interactions (i.e., initial attraction via bound MTB, close range hydrodynamic interactions and MTB exchange, swarm merging, etc.), as well as MTB dispersion within and ejection from swarms, and how to control these phenomena. Combined magnetic/illumination formation/motion control seems promising \cite{bacteria-polygonal-spinning-motion-light-induced}, and one could potentially use seed particles (perhaps also controllable via MF or electric field) in suspension to promote swarm formation where desired. We are also considering an extension of the swarm motion model proposed herein, such that also accounts for swarm drift as observed here and for MTB in \cite{mtb-tunable-hydrodynamics, mtb-tunable-self-assembly-swarms}. Finally, it makes sense to perform more systematic measurements of torque generated by MSR-1, as well as other frequently used MTB, such as AMB-1, and report them for inclusion in a unified database \cite{swimming-organism-data-bank}.

\section*{Acknowledgments}

Klaas Bente (K.B.), Damien Faivre (D.F.), Guntars Kitenbergs (G.K.) and Andrejs Cebers (A.C.) acknowledge the funding from the  "Stochastic dynamics of magnetotactic bacteria at random switching of rotary motors" project of the Baltic-German University Liaison Office, supported by the German Academic Exchange Service (DAAD), with funds from the Foreign Office of the Federal Republic Germany, which initiated this research. G.K. received funding from the PostDocLatvia project no. 1.1.1.2/VIAA/1/16/197. Mihails Birjukovs (M.B.), G.K. and A.C. were funded via the Latvian Council of Science, project A4Mswim, project No. lzp-2021/1-0470. The authors were also supported through the French-Latvian bilateral program "Osmose", project No. LV-FR/2023/3. The authors would like to express gratitude to Mihails Belovs (University of Latvia) for productive discussions regarding the theoretical model for MTB swarm motion, as well as to Mara Smite (University of Latvia) for discussions on MTB swarm formation.

\subsection*{Data \& materials availability}

Experimental data is available on reasonable request, please contact the corresponding author.

\subsection*{Code availability}

Image/data analysis code is available on Image and data analysis code is available on \textit{GitHub}: \href{https://github.com/Mihails-Birjukovs/MTB_swarm_motion_image_data_analysis}{Mihails-Birjukovs/MTB\_swarm\_motion\_image\_data\_analysis
}.

\printbibliography[title={References}]

\begin{appendices}

\section{Velocity components of the swarm-induced induced flow}
\label{secA1}

Relations $v_x(x,y,R)$ and $v_y(x,y,R)$ used in (\ref{eq:velocity-correction}) and (\ref{eq:velocity-expression-final}) -- note that $(x,y,R)$ are scaled by $h$, and velocity by $\Delta v/(2\pi)$.

\begin{eqnarray}
v_{x}(x,y,R)=\arctan(R-x,1-y)-\arctan(-(R+x),1-y)\\ \nonumber
-\arctan(R-x,1+y)+\arctan(-(R+x),1+y)\\ \nonumber
-2y(R-x)/(R-x)^2+(1+y)^2)
-2y(R+x)/\left((R+x)^{2}+(1+y)^2\right)
\end{eqnarray}

\begin{eqnarray}
v_{y}(x,y,R)=1/2\cdot\big(
\ln\left((R-x)^2+(1+y)^2\right)
-\ln\left((R+x)^2+(1-y)^2\right)\\ \nonumber
-\ln\left((R-x)^2+(1+y)^2\right)
+\ln\left((R+x)^2+(1+y\right)^2 \big)\\ \nonumber
+2y(1+y)/((R-x)^2+(1+y)^2)-2y(1+y)/((R+x)^2+(1+y)^2)
\end{eqnarray}

\section{MTB sample preparation}
\label{secA2}

The cultivation medium reported by Heyen and Sch\"uler \cite{Heyen2003} was used to grow the MSR-1 strain. MTB suspension with an optical density of $0.2$ (\textit{NanoPhotometer}\textsuperscript{TM} Pearl at $565~ nm$) was used to form MSR-1 swarms. The medium with the bacteria was degassed using nitrogen for 10 minutes and then inserted into a
rectangular micro-capillary (\# 3520-050, \textit{Vitrocubes}, $T=200~\mu m$) by capillary forces. The loaded micro-capillary was sealed at one end and left open at the other. The resulting oxygen gradient introduces a position with preferred oxygen concentration, where motile bacteria are concentrated \cite{AdlerScience}. This process was accelerated by introducing a magnetic field antiparallel to the oxygen gradient to guide the swimming direction of the bacteria, which corresponds to an ideal case of magneto-aerotaxis in the environment \cite{FaivreSchuler}. After 30 minutes, the concentration of bacteria at this position was high enough to start the swarm experiments.

\section{Detecting and tracking MTB swarm cores}
\label{secA3}

A detailed description and illustration of the underlying image processing methods will be provided in the follow-up methods-focused paper. Here, we provide a brief overview of the developed approach.

\subsection*{Large swarms}

\begin{figure}[htbp]
\centering
\includegraphics[width=0.8\textwidth]{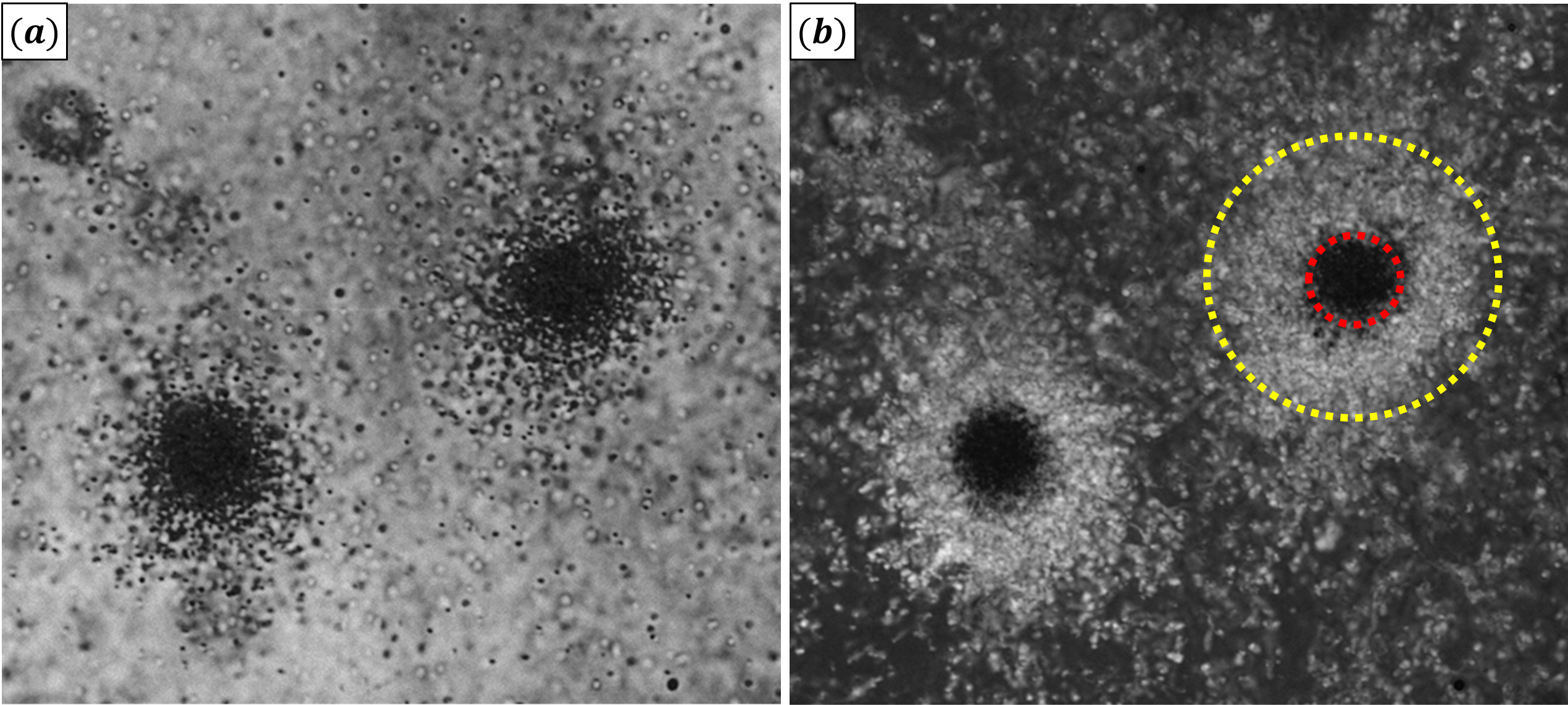}
\caption{(a) A raw image with 2 MTB swarms in the FOV and (b) a pixel-wise luminance standard deviation projection for an image stack, highlighting the swarm structure: the core (red dashed circle) and MTB the "accretion" zone (yellow dashed circle).}
\label{fig:swarm-core-accretion}
\end{figure}

An example of an image with larger-scale MTB swarms in the field of view (FOV) is shown in Figure \ref{fig:swarm-core-accretion}a. Generally, these large swarms have a denser central region where MTB cannot be discerned, i.e. a \textit{swarm core}, surrounded by an \textit{"accretion" zone} where MTB are much sparser, as seen in Figure \ref{fig:swarm-core-accretion}b, which suggests these regions can be isolated and characterized separately.

First, the images are corrected by applying color tone mapping (CTM) \cite{reproduction-of-color-chapter-6}, a reference-less flat field correction (FFC) \cite{wolfram-brightness-equalize} and image inversion (swarm cores are luminance maxima). Noise is filtered using the Gaussian total variation (TV) filter \cite{total-variation-rof-model} and swarm core contrast-to-noise ratio (CNR) is boosted using the multi-pass soft CTM masking (SCTMM) \cite{birjukovs2021resolving, birjukovs-particle-EXIF, birjukovs-solidification-image-processing}. The core regions are segmented using the Otsu method \cite{otsu-thresholding}, and then, for cleanup, one applies Gaussian blurring with repeated Otsu segmentation, morphological erosion (disk kernel) \cite{images-mathematical-morphology}, morphological opening (disk kernel) \cite{images-mathematical-morphology}, filling transform \cite{book-digital-image-processing}, size thresholding, another opening operation, and finally border component removal. The final core segments are approximated by ellipses. While one would ideally expect the swarms to be circular (an optical projection of a cylinder \cite{cebers-ref-11}), realistically they are ellipses.

One then has, at every time stamp, core centroids $\vec{r}(x,y)$, where $x$ and $y$ are image coordinates (horizontal and vertical axis, respectively, with an origin at the lower-left image corner). In this case, nearest-neighbor tracking over pairs of successive frames is sufficient for full trajectory reconstruction.

\subsection*{Small swarms}

\begin{figure}[htbp]
\centering
\includegraphics[width=0.80\textwidth]{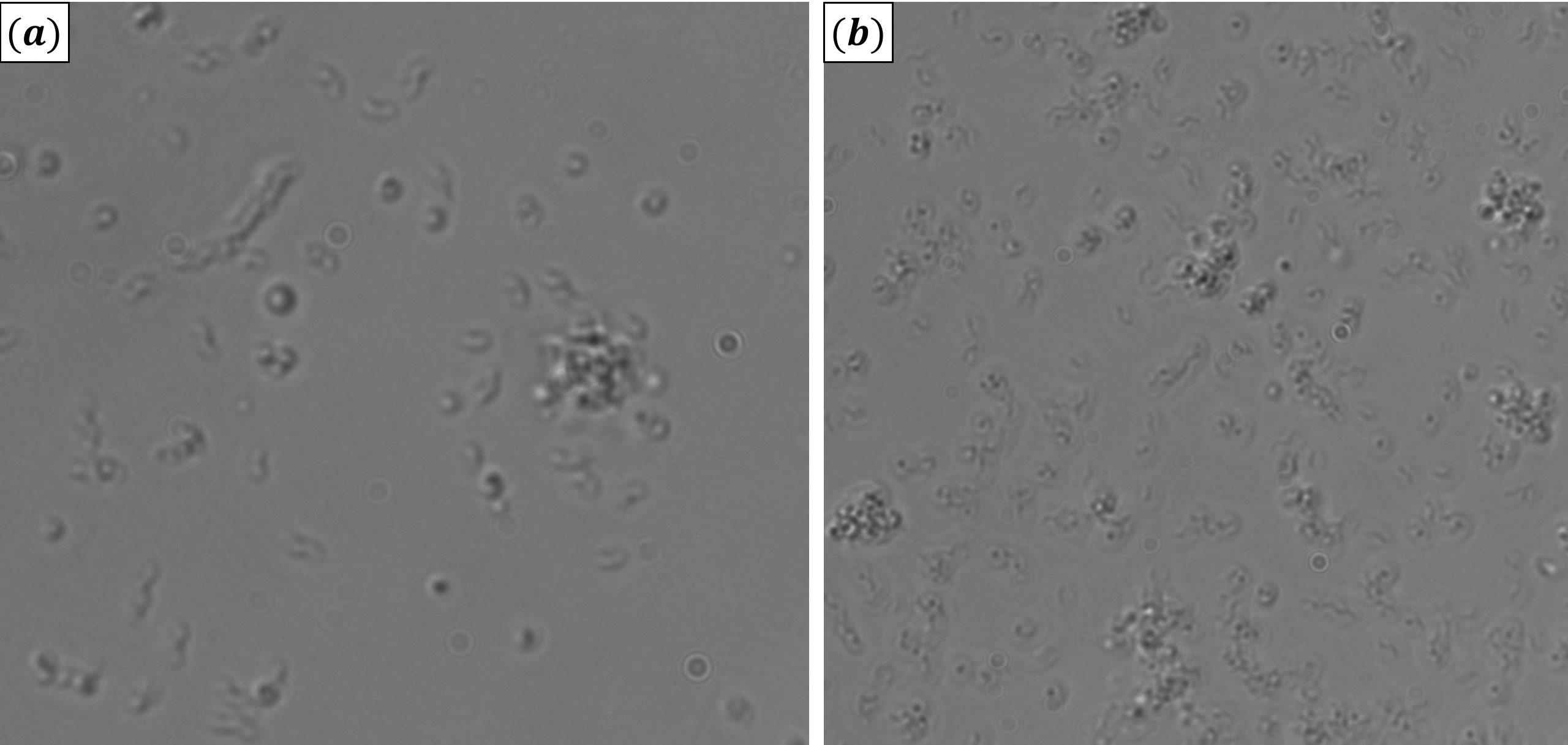}
\caption{Two examples of raw images with small MTB: (a) one and (b) several MTB in the FOV.}
\label{fig:small-swarm-examples}
\end{figure}

Two representative examples of smaller swarms (i.e. less MTB within the core, assuming the classification introduced in Figure \ref{fig:swarm-core-accretion}b), are shown in Figure \ref{fig:small-swarm-examples}.

The key differences between this case and the one with large MTB swarms: first, there are two classes of motile MTB -- ones that respond to MF, and ones that do not. The former are mostly seen as black spots, as in the previous case, while the latter are elongated brighter regions. The latter MTB class, while not complying with MF, are alive and interact with other MTB and swarm cores hydrodynamically. Second, there are more persistent artifacts stemming from stuck/dead MTB and sample glass contamination, as well as some optical artifacts present about MTB, which manifest as bright/dark halos. The framework used for larger swarms is therefore somewhat modified.

Static background artifacts and features are removed by mean background image (or means for time intervals of images if the FOV is adjusted during image acquisition) subtraction. Next, an inverted "mirror" version (\textit{bright/dark branches}) of an image sequence is created, and both versions are processed simultaneously. This is because MTB, especially those oriented parallel to the imaging plane, can have both bright and dark features, some due to optical artifacts, and it makes sense to attempt detecting swarm cores based on all of these features. Both versions/branches of the image sequence are filtered via the same sequence of operations. The first two steps are CTM and FFC. The self-snakes curvature flow (SSCF) filter \cite{self-snakes-curvature-flow} is used for noise filtering, and SCTMM is applied. After the above is done for bright/dark branches, they are combined via image addition, and the resulting images are normalized.

Swarm core segmentation is done via double-Otsu hysteresis binarization \cite{book-digital-image-processing}, followed by morphological dilation \cite{images-mathematical-morphology} and closing \cite{images-mathematical-morphology} (both with disk kernels), and then size thresholding, Gaussian blurring with subsequent Otsu binarization, then morphological opening (disk kernel), another instance of size thresholding, and finally border component removal. When the simpler 2-Otsu hysteresis binarization does not produce adequate results and/or fails to detect swarm cores that are visually present (quite rare), Chan-Vese binarization \cite{getreuer-chan-vese, wolfram-chan-vese} is used instead.

As with the larger swarms, once the above is done and cores are segmented in every frame, their trajectories can be reconstructed, and their properties tracked. In this case, however, simple nearest-neighbor tracking is not always enough, so its multiscale (with respect to search radii over distance/time) version is used instead. In addition, a routine was implemented for connecting trajectories in cases where the FOV was moved to a better position during image acquisition.

\clearpage

\section{Estimating MTB number density $n_s$ for swarm cores}
\label{secA4}

A detailed description and illustration of the underlying image processing methods will be provided in the follow-up methods-focused paper. Here, we provide a brief overview of the developed approach.

\subsection*{Large swarms}

The main idea is to define rectangular interrogation windows about MTB swarm cores (\textit{swarm windows}, SWs) and detect MTB outside the core segments, as roughly delimited in Figure \ref{fig:swarm-core-accretion}b. SWs are defined as in \cite{birjukovs2021resolving} and are based on the core segment size/shape. A local filtering/segmentation methodology developed in \cite{birjukovs-particle-EXIF}, based on a lattice of partially overlapping interrogation windows (IWs) generated for every SW, is then applied to detect MTB within SWs. An algorithm for false positive elimination based on luminance of the segment regions in images, also explained and motivated in \cite{birjukovs-particle-EXIF}, is applied to filtered IWs and the resulting MTB segments. The MTB masks for SWs are then assembled by adding the respective IW MTB masks as in \cite{birjukovs-particle-EXIF}. To further clean up the MTB masks, another instance of false positive elimination, this time for SWs, in performed, and MTB segments that lie within the swarm core region are removed. The procedure is quite similar to the one used for IWs both here (adapted from \cite{birjukovs-particle-EXIF}) and in \cite{birjukovs-particle-EXIF, birjukovs-particle-track-curvature-stats}.

Two more issues remain. First, since the SWs are rectangular as opposed to the circular/elliptic shapes of swarm cores, the MTB near the corners of SWs will introduce a bias into MTB number density distributions, and therefore must be removed. Second, some of the remaining MTB segments are distinctly irregular and acircular -- the MTB of interest that respond to MF are oriented normally to the imaging plane and should present in the FOV as roughly circular dark objects, and other objects must be eliminated.

The first problem is solved using cutoff masks formed by applying morphological dilation \cite{images-mathematical-morphology} to the swarm core segments present in SWs, which conformally inflates the swarm core segment until it reaches the nearest SW boundary. The MTB segments outside this inflated mask (i.e., near image corners) are removed. The second problem is solved by introducing a geometric criterion for MTB based on aspect ratio and circularity. The final output in the form of MTB coordinates now enables one to compute MTB radial $n_s$ distribution for each swarm over time. The maximum of $n_s$ at the swarm core boundary is taken as the core $n_s$ value, since it is assumed that cores are single MTB layers and $n_s$ reached its saturation value at the swarm boundary.

\subsection*{Small swarms}

The same IW-based approach used for large swarms is applied here. However, as with core segmentation, two branches of SW images are generated and processed separately. The SW (global) false positive filtering is performed, as with large swarms, for bright and dark branches. Once the processing is done, both branches are combined and then filtered further using geometric and distance criteria for MTB to eliminate optical imaging artifacts while preserving MTB segments. The geometric criterion here is based on caliper width to equivalent disk diameter ratio, convex hull area, aspect ratio and circularity. Once the final centroids are obtained for every MTB swarm in every frame, one can derive $n_s$.

\section{Data analysis for Figure \ref{fig:experiment-swarm-motion-direction-2}}
\label{secA5}

Velocity time series with uncertainty bands as shown in Figures \ref{fig:experiment-swarm-motion-direction-2}b,c are obtained from raw trajectory data as follows. First, tracks are filtered with a median filter (1 point wide kernel) and a mean filter (3 point wide kernel), and then velocity component time series were computed. The latter were then filtered for outliers using quantile spline envelopes (QSE) \cite{antonov-qse} (using the \textit{Wolfram Mathematica} code package available on \textit{GitHub}: \href{https://github.com/antononcube/MathematicaForPrediction/blob/master/QuantileRegression.m}{Anton Antonov
(\textit{antononcube}): MathematicaForPrediction/QuantileRegression.m}) via a procedure outlined in \cite{birjukovs2021resolving} (except its last step). The time series were then binned (bin size based on a percentage of the time series length) and bin means and standard deviations computed. For all swarm trajectories considered here, the parameters were as follows (please refer to \cite{birjukovs2021resolving}): $q=0.95$ with 3-rd order splines and $\nint{5\% \cdot N}$ spline knots for QSEs ($N$ is the number of points in the dataset); $N_\text{QSE} = \nint{12.5\% \cdot N}$ and $\delta_\text{b} = \nint{0.75\% \cdot N}$ (point density-adaptive physical bin size); the TV regularization parameter \cite{total-variation-rof-model} for QSEs was $0.5$.

\end{appendices}

\end{document}